\documentclass[a4paper,UKenglish,cleveref, autoref, thm-restate]{lipics-v2019}

\usepackage{tikz-cd}
\usepackage{amsmath} 
\usepackage{hyperref}
\usepackage{tikz}

\usetikzlibrary{calc,positioning,shapes,shadows,arrows,fit}
\usetikzlibrary{arrows.meta}

 \DeclareFontFamily{OMX}{MnSymbolE}{}
\DeclareFontShape{OMX}{MnSymbolE}{m}{n}{
    <-6>  MnSymbolE5
   <6-7>  MnSymbolE6
   <7-8>  MnSymbolE7
   <8-9>  MnSymbolE8
   <9-10> MnSymbolE9
  <10-12> MnSymbolE10
  <12->   MnSymbolE12}{}
\DeclareSymbolFont{mnlargesymbols}{OMX}{MnSymbolE}{m}{n}
\SetSymbolFont{mnlargesymbols}{bold}{OMX}{MnSymbolE}{b}{n}
\DeclareMathDelimiter{\llangle}{\mathopen}{mnlargesymbols}{'164}{mnlargesymbols}{'164}
\DeclareMathDelimiter{\rrangle}{\mathclose}{mnlargesymbols}{'171}{mnlargesymbols}{'171}

\newtheorem{observation}{Observation}

\usepackage{thm-restate}
\usepackage{hyperref}

\usepackage{cleveref}

\newcommand{\ex}{{X}}

\renewcommand{\u}{\boldsymbol{u}}
\renewcommand{\v}{\boldsymbol{v}}

\newcommand{\F}{\mathbb{F}}
\newcommand{\N}{\mathbb{N}}

\newcommand{\M}{\mathcal{M}}

\renewcommand{\angle}[1]{{\langle} #1 {\rangle}}
\newcommand{\dangle}[1]{{\llangle} #1 {\rrangle}}

\newcommand{\FX}{\F\angle{\ex}}

\newcommand{\FM}{\F\angle{M}}

\DeclareMathOperator{\poly}{\mbox{\small\rm poly}}

\title{Equivalence Testing of Weighted Automata over Partially
  Commutative Monoids} 

\titlerunning{Equivalence Testing of Weighted Automata} 

\author{V. Arvind}{Institute of Mathematical Sciences (HBNI), Chennai, India}{email: arvind@imsc.res.in}{}{} 
\author{Abhranil Chatterjee}{Institute of Mathematical Sciences (HBNI), Chennai, India}{email: abhranilc@imsc.res.in}{}{} 
\author {Rajit Datta}{Chennai Mathematical Institute, Chennai, India}{email: rajit@cmi.ac.in}{}{}
\author {Partha Mukhopadhyay}{Chennai Mathematical Institute, Chennai, India}{email: partham@cmi.ac.in}{}{}

\authorrunning{V.Arvind, A.Chatterjee, R.Datta, and P.Mukhopadhyay} 

\Copyright{V.Arvind, A.Chatterjee, R.Datta, and P.Mukhopadhyay} 



\category{} 

\relatedversion{} 

\supplement{}

\nolinenumbers 

\EventEditors{John Q. Open and Joan R. Access}
\EventNoEds{2}
\EventLongTitle{42nd Conference on Very Important Topics (CVIT 2016)}
\EventShortTitle{CVIT 2016}
\EventAcronym{CVIT}
\EventYear{2016}
\EventDate{December 24--27, 2016}
\EventLocation{Little Whinging, United Kingdom}
\EventLogo{}
\SeriesVolume{42}
\ArticleNo{23}

\begin{document}

\maketitle

\begin{abstract}
  We study the \emph{equivalence} testing of automata over partially commutative monoids (pc monoids) and show efficient algorithms in special cases, exploiting the structure of the underlying non-commutation graph 
  of the monoid.
  
  Specifically, if the clique edge cover number of the non-commutation graph of the pc
  monoid is a constant, we obtain a deterministic quasi-polynomial
  time algorithm.  As a consequence, we also obtain the first deterministic quasi-polynomial time algorithms for equivalence testing of $k$-tape weighted automata and for equivalence testing of deterministic $k$-tape automata for constant $k$. Prior to this, a randomized polynomial-time algorithm for the above problems was shown by Worrell~\cite{Worrell13}.
  
  We also consider pc monoids for which the
  non-commutation graphs have cover consisting of at most $k$ cliques
  and star graphs for any constant $k$. We obtain randomized
  polynomial-time algorithm for equivalence testing of
  weighted automata over such monoids. 
  
  Our results are obtained by designing efficient zero testing algorithms for weighted automata over such pc monoids.

\end{abstract}
\newpage

\section{Introduction}\label{intro}

Testing equivalence of multi-tape finite automata is a fundamental
problem in automata theory. For a $k$-tape automaton, we usually
denote by $\Sigma_1, \ldots, \Sigma_k$ the mutually disjoint alphabets
for the $k$ tapes, and the automaton accepts a subset of the product
monoid $\Sigma^*_1\times\cdots\times \Sigma^*_k$. Two multi-tape
automata are equivalent if they accept the same subset. It is
well-known that equivalence testing of multi-tape
\emph{non-deterministic} automata is undecidable~\cite{Griff68}.

For 2-tape \emph{deterministic} automata equivalence testing was shown
to be decidable in the 1970's
\cite{Bird73,Val74}. In \cite{Beeri76} an exponential upper bound was shown. Subsequently, a polynomial-time algorithm
was obtained~\cite{FG82} and the authors conjectured that equivalence
testing of deterministic $k$-tape automata for any constant $k$ is
in polynomial time.


A closely related problem is testing the \emph{multiplicity
  equivalence} of multi-tape automata.  Intuitively, the
multiplicity equivalence testing problem is to decide whether for each
tuple in the product monoid $\Sigma^*_1\times\cdots\times \Sigma^*_k$,
the number of accepting paths in the two input automata are the same.
Since a deterministic automaton has at most one accepting path for
each word, equivalence testing of two deterministic $k$-tape automata
coincides with multiplicity equivalence testing. More generally, for
weighted automata, equivalence testing is to decide if
the coefficient of each word (over a field or ring) is the same in the
given automata.
For the weighted case, the equivalence testing is in deterministic polynomial time
for one-tape automata~\cite{Sch61,Tze92}. Such an algorithm for the
$k$-tape case remained elusive for a long time. Eventually the 
equivalence testing of $k$-tape non-deterministic weighted automata was shown
\emph{decidable} by Harju and Karhum\"{a}ki~\cite{HK91} using the
theory of free groups~\footnote{They were also the first to settle the decidability of equivalence problem for deterministic multi-tape automata. }. No nice complexity-theoretic upper bound was
known, until recently Worrell~\cite{Worrell13} obtained a
\emph{randomized} polynomial-time algorithm for testing the
 equivalence of $k$-tape weighted non-deterministic automata (and
equivalence testing of deterministic $k$-tape automata) for any
constant $k$. Worrell takes a different approach via Polynomial
Identity Testing (PIT). In~\cite{Worrell13}, Worrell explicitly
raised the problem of finding an efficient \emph{deterministic}
algorithm for equivalence problem for $k$-tape weighted automata
for any fixed $k$.

In this paper, we show that the equivalence testing for
$k$-tape weighted automata can be solved in \emph{deterministic}
quasi-polynomial time. This immediately yields the first deterministic
quasi-polynomial time algorithm to check the equivalence of
deterministic $k$-tape automata, making progress on a question asked
earlier~\cite{FG82, HK91}. In fact, our proof technique shows a
stronger result that we explain now. The product monoid
$M=\Sigma^*_1\times\cdots\times \Sigma^*_k$ associated with $k$-tape
automata is a \emph{partially commutative monoid} (henceforth \emph{pc monoid}), in the sense that
any two variables $x\in\Sigma_i, y\in\Sigma_j, i\ne j$ commute with
each other whereas the variables in the same tape alphabet $\Sigma_i$
are mutually non-commuting. We associate a \emph{non-commutation
  graph} $G_M$ with $M$ to describe the non-commutation relations:
$(x,y)$ is an edge if and only if $x$ and $y$ do not commute. 
If there is no edge $(x,y)$ in $G_M$, the words $xy$ and $yx$ are considered to be equivalent as $x$ and $y$ commute. 
The notion of words over a pc monoid and equivalence of two words are discussed in details in Section~\ref{main-section}.
For the
$k$-tape case, the vertex set of $G_M$ is $\Sigma_1 \cup \ldots
\cup\Sigma_k$ and $G_M$ is clearly the union of $k$ disjoint cliques,
induced by each $\Sigma_i$, forming a \emph{clique edge cover} of size $k$. 
For convenience, each isolated vertex is a clique of size one. 

In this paper, we obtain an equivalence testing algorithm for weighted automata 
over any pc monoid whose non-commutation graph has
a constant size clique edge cover (\emph{not necessarily} disjoint, and all the isolated vertices are part of the cover). In
short, we call such monoids as \emph{$k$-clique monoids} where the
clique edge cover size is bounded by $k$. Since two weighted automata $A$ and 
$B$ are equivalent if and only if the difference automaton $C=A-B$ is a zero weighted automaton, 
we prefer to describe the results in terms of zero testing of weighted automaton. Here the difference of two weighted automata has an obvious 
meaning: the weight of each word $w$ in $C$ is the difference between the weights of $w$ in $A$ and $B$. 
The words over any pc monoid are defined with respect to the equivalence relation induced by the 
non-commutation graph of the pc monoid. This is explained in Section \ref{main-section}.  
Let $\F$ be an infinite field from where the weights are taken.   

\begin{theorem}\label{main-theorem-1}
  Let $A$ be a given $\F$-weighted automaton of size $s$ over a
  pc monoid $M$ for which the non-commutation graph
  $G_M$ has a clique edge cover of size $k$. Then, the zero  testing  of $A$ can be decided in deterministic
  $(nks)^{O(k^2\log ns)}$ time. Here $n$ is the size of the alphabet of
  $M$, and the clique edge cover is given as part of the input.
\end{theorem}

As an immediate corollary, the above theorem yields a deterministic quasi-polynomial
time algorithm for equivalence testing of $k$-tape weighted automata (also
for equivalence testing of deterministic $k$-tape automata). Notice
that, for the $k$-tape case, the clique edge cover of size $k$ is also part
of the input since for each $1\leq i \leq k$, the $i^{th}$ tape alphabet
$\Sigma_i$ is explicitly given.

Next we address equivalence testing over more general pc monoids $M$. 
$M$ is a \emph{$k$-monoid} if its
non-commutation graph $G_M$ is a union $G_1\cup G_2$ of two graphs,
where $G_1$ has a clique edge cover of size at most $k'$ and $G_2$ has a
vertex cover of size at most $k-k'$ (hence the edges of $G_2$ can be covered by
$k-k'$ many star graphs).  We show that equivalence
testing over $k$-monoids has a randomized polynomial-time
algorithm. One can also see this result as a generalization of Worrell's result~\cite{Worrell13}.
 

\begin{theorem}\label{main-theorem-2}
  Let $A$ be a given $\F$-weighted automaton of size $s$ over a
  $k$-monoid $M$. Then the zero testing of $A$ can be decided in
  randomized $(ns)^{O(k)}$ time. Here $n$ is the size of the alphabet
  of $M$.
\end{theorem}

\begin{remark}
What is the complexity of equivalence testing for weighted automata over
general pc monoids? The non-commutation graph $G_M$
of any pc monoid $M=(\ex^*,I)$ has a clique edge 
covering of size bounded by $|{X}|\choose 2$. Hence, the
above results give an exponential-time algorithm. Note that if
$G_M$ has an \emph{induced matching} of size more than $k$ then
$M$ is not a $k$-monoid. Call $M$ a \emph{matching monoid}
if $G_M$ is a perfect matching. It follows from Lemma~\ref{lemma-equivalence-pcmonoid-to-partitioned} shown in Section~\ref{main-section}, that equivalence testing over arbitrary pc monoids is deterministic polynomial-time reducible to equivalence testing over matching monoids (if $G_M$ has isolated vertices, one can add a new vertex (variable) for each isolated vertex and introduce a matching edge between them). 
\end{remark}

Various automata-theoretic problems have been studied in the setting
of pc monoids. For example, pc
monoids have found applications in modelling the behaviour of
concurrent systems~\cite{Maz86}. Droste and Gastin~\cite{DG97} have
studied the relation between recognizability and rationality over
pc monoids. Broadly, it is interesting to
understand and identify the results in algebraic automata theory that
can be generalized to the setting of pc monoids.
 
\vspace{0.15 cm}
\noindent\textbf{Proof Overview :} 
Now we briefly discuss the main ideas behind our results. Worrell's key 
insight \cite{Worrell13} is to reduce $k$-tape automata equivalence problem to a suitable instance of polynomial identity testing over non-commuting variables, which can be solved in randomized polynomial time \cite{AL50,BW05, MVV87}. Our strategy too is to carry out reductions to polynomial identity testing problem. Since we are considering automata over general pc monoids and we aim to design efficient deterministic algorithms, we require additional ideas. First, we suitably apply a classical algebraic framework to transfer the zero testing problem over general pc monoids to pc monoids whose non-commutation graphs are disjoint union of
cliques \cite{CLR90, Diekert90}. This allows us to prove a Sch\"{u}tzenberger~\cite{Sch61}
type theorem over \emph{general pc
  monoids} which says that any nonzero weighted automata of size $s$
over any pc monoid, must have a nonzero word within
length $\poly(s,n)$ where $n$ is the size of the alphabet. Furthermore, this also allows us to reduce the zero testing of weighted automata to polynomial identity testing for algebraic branching programs over pc monoids. It turns out that the latter problem can be solved by suitably adapting a black-box polynomial identity test for noncommutative algebraic branching programs based on hitting sets due to Forbes and Shpilka~\cite{FS13}. Our algorithm recursively builds on this result, ensuring that the resulting hitting set remains of quasi-polynomial size, like the Forbes-Shpilka hitting set \cite{FS13}. This requires coupling a result of Sch\"{u}tzenberger related to the Hadamard product of weighted automata (\cite{Sak09}, Theorem 3.2) with our algebraic framework. The proof of Theorem~\ref{main-theorem-2} also follows a similar line of argument. First
we give a randomized polynomial-time identity testing algorithm over pc monoids whose non-commutation graph is a star graph. Then a composition lemma yields an identity testing algorithm over $k$-monoids. 


The paper is organized as follows. In Section~\ref{section-preli}, we
give some background. We prove a Sch\"{u}tzenberger type
theorem for automata over pc monoids in
Section~\ref{main-section}. Theorem~\ref{main-theorem-1}
is presented in Section~\ref{pcabppitsec}, and Theorem~\ref{main-theorem-2} in Section~\ref{random-pit}. Some proof details are in the appendix.
\vspace{-0.1 cm}

\section{Preliminaries }\label{section-preli}

We recall some basic definitions and results, mainly from automata theory and arithmetic circuit complexity, and define some notation used in the subsequent sections.

\vspace{0.15 cm}
\noindent\textbf{Notation : }
Let $\F$ be an infinite field. 
Let $\M_t(\F)$ denote the ring of $t \times t$ 
matrices over $\F$. For matrices $A$ and $B$ of sizes $m \times n$ and $p \times q$ respectively, their Tensor (Kronecker) product $A \otimes B$ is defined as $\left(a_{ij}B\right)_{1\leq i\leq m, 1\leq j\leq n}$. The dimension of $A \otimes B$ is $pm\times qn$.  Given bases $\{v_i\}$ and $\{w_j\}$ for the vector spaces $V$ and $W$, the vector space 
$V\otimes W$ is the tensor product space with a basis $\{v_i\otimes w_j\}$. 

For a series (resp. polynomial) $S$ and a word (resp. monomial) $w$, let $[w]S$ denote  the coefficient of $w$ in the series $S$ (resp. polynomial). In this paper, we consider weighted automata over a field $\F$ and
alphabet (or variables) $\ex =\{x_1, \ldots, x_n\}$. 

We also consider coverings of graphs : a graph $G=(\ex,E)$ is said to have a graph covering $\{ G_i = (\ex_i , E_i) \}^{k}_{i=1}$ of size $k$ if $\ex= \cup^{k}_{i=1} \ex_i$ and $E= \cup^{k}_{i=1} E_i$.


 
 
\vspace{0.15 cm}
\noindent\textbf{Automata Theory : }
 We recall some basic definitions from automata theory. More details can be found in the
Berstel-Reutenauer book \cite{BR11}.

Let $K$ be a semiring and $\ex$ be an alphabet. A $K$-weighted
automaton over $\ex$ is a $4$-tuple, ${A} = (Q, I, E, T)$,
where $Q$ is a finite set of states, and the mappings $ I, T : Q \to
K$ are weight functions for entering and leaving a state respectively,
and $E: Q\times {\ex} \times Q \to K$ is the weight of each
transition. We define $|Q|$, the number of states, to be the size of
the automaton.  A path is a sequence of edges : $(q_0, a_1, q_1) (q_1,
a_2, q_2)\cdots(q_{t-1},a_t,q_t)$. The weight of the path is the
product of the weights of the edges. The formal series $S \in
K\dangle{\ex}$ which is the (possibly infinite) sum of the weights
over all the paths is \emph{recognized} by ${A}$. Then, for
each word $w=a_1 a_2 \cdots a_t\in \ex^*$, the contribution of all the
paths for the word $w$ is given by $[w]S = \sum_{q_0, \ldots, q_t \in
  Q} I(q_0)\cdot E(q_0, a_1, q_1) \cdots E(q_{t-1}, a_t, q_t)\cdot T(q_t)$.

A $K$-weighted automaton ${A}$ with
\emph{$\epsilon$-transitions} over $\ex$ is defined with $E$ modified,
such that $E: Q\times \{\ex\cup \epsilon\} \times Q \to K$. Let
$A_0\in \mathbb{M}_{|Q|}(K)$ be the transition matrix for the
$\epsilon$-transitions. An automaton computes a valid formal series in
$K\dangle{\ex}$, if and only if $\sum_k A^k_0$ converges. In that
case, another automaton ${A'}$ without $\epsilon$-transitions
computing the same series can be constructed efficiently \cite{LS12}. 
Henceforth, we consider all automata are valid and free of
$\epsilon$-transitions.
 
The following basic result by Sch\"{u}tzenberger \cite{Sch61} is important 
for the algorithmic results presented in this paper.  

\begin{theorem}[Sch\"{u}tzenberger]\label{szbgr}
Let $K$ be a subring \footnote{For some applications, this could also be subsemirings as originally proved~\cite{Sch61}.} of a division ring and ${A}$ be a
$K$-weighted automaton with $s$ states representing a series $S$ in
$K\dangle{\ex}$. Then $S$ is a nonzero series if and only if there
is a word $w\in \ex^{*}$ of length at most $s-1$, such that $[w]S$ is nonzero.  
\end{theorem}
Now, we recall the definition of weighted \textbf{multi-tape automata} 
following Worrell's work
\cite{Worrell13}.
Let $M$ be the pc monoid over variables $\ex =
\ex_1\cup \cdots \cup \ex_k$ defined as follows: the variables in each
$\ex_i$ are non-commuting, but for all $i\ne j$ and any $x\in\ex_i,
y\in\ex_j$ we have $xy=yx$. As defined already, the transition function 
$E$ is a mapping $Q\times {\ex} \times Q \to K$. A path is a sequence of edges : $(q_0, x_1, q_1) (q_1,
x_2, q_2)\cdots(q_{t-1},x_t,q_t)$ where each $x_i\in\ex_j$ for some $j$. The label of the run is 
$m=x_1 x_2 \cdots x_t$ in the pc monoid $M$, and $[m]\mathcal{A}$ is the total contribution of all the runs having the label equivalent to $m$. 
 
An automaton is \emph{deterministic} if the set of states can be
partitioned as $Q = Q^{(1)} \cup \ldots \cup Q^{(k)}$, where states in
$Q^{(i)}$ read input only from the set $\ex_i$ which is the alphabet of $i^{th}$ tape, and each state has a
single transition for every input variable. Thus, a deterministic
automaton has at most one accepting path for each input $m\in M$.

\vspace{0.15 cm} 
\noindent\textbf{Arithmetic Circuit Complexity : } An \emph{algebraic branching program} (ABP) is a directed acyclic
graph with one in-degree-$0$ vertex called \emph{source}, and one
out-degree-$0$ vertex called \emph{sink}. The vertex set of the graph
is partitioned into layers $0,1,\ldots,\ell$, with directed edges 
only between adjacent layers ($i$ to $i+1$). The source and the sink
are at layers zero and $\ell$ respectively. Each edge is labeled by an
affine linear form over $\F$. The polynomial computed by the ABP is
the sum over all source-to-sink directed paths of the product of
linear forms that label the edges of the path. The maximum number of
nodes in any layer is called the width of the algebraic branching
program. The size of the branching program is taken to be the total number of nodes.  

Equivalently, the computation of an algebraic branching program can be
defined via the iterated matrix product $\u^T M_1 M_2 \cdots M_{\ell} \v$, where $\u,\v$ are vectors in $\F^w$ and
each $M_i$ is a $w \times w$ matrix whose entries are affine linear
forms over ${\ex}$. Here $w$ corresponds to the ABP width and $\ell$
corresponds to the number of layers in the ABP. If $\ex$ is a set of
non-commuting variables then the ABP is a noncommutative algebraic
branching program (e.g., see \cite{Ni91}). 

Now we recall some results from noncommutative polynomial identity testing. 
Let $S\subset \FX$ be a subset of polynomials in the 
noncommutative polynomial ring $\FX$ where $\ex = \{x_1,\ldots,x_n\}$. Given a
mapping $v:\ex\to \M_t(\F)$ from variables to $t\times t$ matrices, it
defines an \emph{evaluation map} defined for any polynomial $f\in\FX$
as $v(f) = f(v(x_1),\ldots,v(x_n))$. A collection $H$ of such
evaluation maps is a \emph{hitting set} for $S$, if for every nonzero
$f$ in $S$, there is an evaluation $v\in H$ such that $v(f)\ne 0$.

Let $S_{n,d,s}$ denote the subset of polynomials $f$ in
$\F\angle{\ex}$ such that $f$ has an algebraic branching program of
size $s$ and $d$ layers. Forbes and Shpilka~\cite{FS13} have shown that a hitting set $H_{n,d,s}$ of quasi-polynomial size for
$S_{n,d,s}$ can be constructed in quasi-polynomial time.

\begin{theorem}[Forbes-Shpilka]\label{forbesshpilka}
  For all $s,d,n \in \N$ if $|\F|\geq \poly(d,n,s)$, then there is a
  set $H_{n,d,s}$ which is a hitting set for $S_{n,d,s}$. Further $|
  H_{n,d,s}| \leq (sdn)^{O(\log d)}$ and there is a deterministic
  algorithm to output the set $H_{n,d,s}$ in time $(sdn)^{O(\log d)}$.
\end{theorem}

\section{A Sch\"{u}tzenberger Type Theorem for Partially Commutative Monoids}\label{main-section}

In this section, we prove a theorem in the spirit of Theorem \ref{szbgr} over general pc monoids.  

\vspace{0.12 cm} 
\noindent\textbf{Pc monoids and associated partitioned pc monoids :}
Let $\ex$ be a finite alphabet (equivalently, variable set). A pc monoid $M$ over $\ex$ is usually 
denoted as $M=(\ex^*, I)$ where $I\subseteq \ex\times\ex$ is a symmetric and reflexive binary relation such that 
$(x_1, x_2)\in I$ if and only if $x_1 x_2 = x_2 x_1$ in $M$. Let $\tilde{I}$ be the \emph{congruence} generated by $I$ using the 
transitive closure. The monoid elements are defined as the congruence classes $\tilde{m}$ for $m\in\ex^*$. In other words, 
$M$ is a factor monoid of $\ex^*$ generated by $\tilde{I}$.  
The \emph{non-commutation graph} $G_M = (\ex, E)$ of $M$ is a simple
undirected graph such that $(x_1,x_2)\in E$ if and only if 
$(x_1,x_2)\notin I$. 

A $k$-partitioned pc monoid is a pc monoid for which the non-commutation graph can be partitioned into $k$ vertex-disjoint 
subgraphs. Given any pc monoid $M$, we can associate a partitioned pc monoid $M'$ with it, such that $M$ is isomorphic to 
a submonoid of $M'$, as follows. Let $\{G_i\}_{i=1}^k$ be the $k$-cover for $G_M$ where $G_i = (\ex_i, E_i)$. Consider a set of 
variables $\widehat{\ex} = \{x_{t i} : 1\leq t\leq n, 1\leq i\leq k\}$. Do a new labelling of the graph $G_i$ by changing the variable 
$x_t\in\ex_i$ by $x_{t i}$. In this process obtain the graphs $G'_1, \ldots, G'_k$ which are vertex disjoint. The edges in $G'_i$ are naturally 
induced by $G_i$. For each $1\leq i\leq k$, 
the new pc monoid $M'_i$ has $G'_i$ as its non-commutation graph. Finally, $M'$ be the pc monoid generated by $M'_1, \ldots, M'_k$ 
and the alphabet $\ex' = \cup_{i=1}^k V(G'_i)$. By construction, the non-commutation graph $G_{M'}$ is the disjoint union of 
$G'_1, \ldots, G'_k$. As $\F$-algebra $\F\angle{M'}$ is isomorphic to the tensor product of the $\F$-algebras $\F\angle{M'_1}\otimes \cdots \otimes \F\angle{M'_k}$.  

It is a classical result that $M$ is isomorphic to a submonoid of $M'$ \cite{CLR90, Diekert90, DLM06} via the map $\psi$, which we define next.    

\begin{lemma}\label{main-iso-lemma}
Let $\psi : \F\angle{M}\to \F\angle{M'}$ be the map such that $\psi(m) = m_1\otimes m_2\otimes\cdots\otimes m_k$ for any monomial $m$ in $M$ and extend by linearity. Here for $1\leq i\leq k$, the monomial $m_i$ is obtained from the part of $m$ (after erasing the letters not in $\ex_i$) by labelling 
$x_t$ in $\ex_i$ by $x_{t i}$. Then, $\psi$ is an injective homomorphism. 
\end{lemma}

\begin{remark}\label{classical-map}
To fit with our notation, we include a self-contained proof in the appendix.  
\end{remark}

Using Lemma \ref{main-iso-lemma}, we can show that the zero testing for weighted automata over pc monoids reduces to zero testing 
of weighted automata over partitioned pc monoids in deterministic polynomial time. More formally, we show the following result. 

\begin{lemma}\label{lemma-equivalence-pcmonoid-to-partitioned}
  Let $A$ be the given $\F$-weighted automaton of size $s$ over a pc monoid $M$, for which the non-commutation
  graph $G_M$ has $k$-covering $\{G_i = (\ex_i,
  E_i)\}_{i=1}^k$. Then the zero testing of $A$ reduces to the zero testing of another $\F$-weighted automaton $B$ over the associated 
  partitioned pc monoid $M'$ in deterministic polynomial time. Moreover the size of the automaton $B$ is $\poly(n,s,k)$. 
 \end{lemma} 
 
 \begin{proof}
The automaton $B$ is simply obtained by applying the map $\psi$ on the variables in $M$. For a variable $x_t$, let $J_t\subseteq \{1,2,\ldots,k\}$ be the set of indices 
such that, $i\in J_t$ if and only if $x_t\in \ex_i$. Then $\psi(x_t) = \eta_{i_1}\otimes \cdots \otimes \eta_{i_{|J_t|}}$ where $i_1<i_2<\cdots<i_{|J_t|}$ and for each $j$, $i_j\in J_t$.  
Now for each $q_0,q_k\in Q$ such that $(q_0,x_t,q_k)\in E$ and $wt(q_0,x_t,q_k) = \alpha\in \F$, we introduce new states $q_1,\ldots,q_{|J_t|-1}$ and for each $j\leq |J_t|-1$, add 
the edge  $e_j = (q_{j-1},\eta_{i_j},q_j)$ in $E$ and $wt(e_1) = \alpha$ and for other newly added edges the weight is $1$. 
Since the number of edges in $A$ is $O(n s^2)$, it is easy to see the number of nodes in $B$ is $O(n s^2 k)$. The fact that $A$ is zero if and only if $B$ is zero follows from Lemma \ref{main-iso-lemma}.  
\end{proof} 

Worrell has already proved that the zero testing of weighted automata over partitioned monoids whose non-commutation graphs are the union of disjoint cliques, can be reduced to the identity testing of noncommuatative ABPs \cite{Worrell13}. We restate the following proposition from Worrell's paper in a form that fits with our framework. 

\begin{proposition}[Adaptation of Proposition 5 of
  \cite{Worrell13}]\label{worrell-prop}
 Let $A$ be a given $\F$-weighted automaton of size $s$ over a partitioned pc monoid $M$ computing a series $S$. Moreover the non-commutation 
 graph $G_M$ is the disjoint union of $k$ cliques. Let $N$ be the transition matrix of $A$. Then $S$ is a zero series if and only if  
 the ABPs $\boldsymbol{u}^T N^\ell \boldsymbol{v} = 0$ for each $0\leq \ell \leq s - 1$, where $u,v$ are vectors in $\F^s$. 
 \end{proposition}
Combining Lemma~\ref{lemma-equivalence-pcmonoid-to-partitioned} and Proposition~\ref{worrell-prop} we obtain the following generalization of
Sch\"{u}tzenberger's theorem~\cite{Sch61} over arbitrary pc monoids.   

\begin{theorem}[A Sch\"{u}tzenberger type theorem]\label{equivalence-to-pit}
  Let $A$ be a given $\F$-weighted automaton of size $s$ over
  any pc monoid $M$ representing a series $S$. Then $S$ is a nonzero series if and only if 
  there exists a word $w\in \ex^*$ such that $[w]S$ is nonzero and the length of $w$ 
  is bounded by $O(n^3 s^2)$. 
\end{theorem}
\begin{proof}
Observe that the non-commutation graph $G_M$ has a trivial clique edge cover of size $\leq n^2$ where $n$ is the size of the alphabet. Then we apply 
Lemma \ref{lemma-equivalence-pcmonoid-to-partitioned} to conclude that $S$ is a zero series if and only if the series $S'$ computed by the $\F$-weighted automaton 
$B$ over the associated partitioned pc monoid (whose non-commutation graph is a disjoint union of cliques) is zero. The size $s'$ of $B$ is bounded by $O(n^3 s^2)$. Now we use 
Proposition \ref{worrell-prop} to see that $S'$ is identically zero if and only if the ABPs $\boldsymbol{u}^T N^\ell \boldsymbol{v} = 0$ for each $0\leq \ell \leq s' - 1$ are identically zero 
where $N$ is the transition matrix of $B$. Now notice that under the image of $\psi$ map, the length of any word can only increase. In other words, for any word $w : |\psi(w)|\geq |w|$. 
Using this, we conclude that $(S'=\psi(S))^{\leq s'-1}$ is a nonzero polynomial. Since $\psi$ is injective, it must be the case that $S^{\leq s'-1}$ is also a nonzero polynomial and the proof of the theorem follows.  
\end{proof}

\section{Deterministic Algorithm for Zero Testing  of Weighted Automata Over $k$-Clique
  Monoids}\label{pcabppitsec}

Recall from Section \ref{intro}, that a \emph{$k$-clique monoid} is a pc monoid $M$ whose
non-commutation graph $G_M$ has a clique edge cover of size $k$. In this
section, we show that the zero testing problem for automata over
$k$-clique monoids for constant $k$ can be solved in deterministic
quasi-polynomial time. In fact, using Lemma~\ref{lemma-equivalence-pcmonoid-to-partitioned} and Proposition~\ref{worrell-prop}, it is straightforward to observe that 
the zero testing problem reduces to the polynomial identity testing of ABPs over partitioned pc monoids whose non-commutation graph is a disjoint union of $k$ cliques. Thus the main purpose of this section is to develop identity testing algorithm for ABPs computing polynomials in $\F\angle{\ex_1}\otimes \cdots \otimes \F\angle{\ex_k}$, where each set 
$\ex_j=\{x_{ij}\}_{1\leq i\leq n}$ is of size $n$, and the sets are mutually disjoint. The parameter $k$ is a constant. This will suffice to prove Theorem \ref{main-theorem-1}. 

We first formally define the concept of evaluation and partial evaluation of polynomials over algebra. 

\vspace{0.18 cm}
\noindent\textbf{Evaluation of a polynomial over algebras :}  
Given a polynomial $f\in
\F\angle{\ex_1}\otimes\cdots\otimes\F\angle{\ex_k}$ and a $k$-tuple of $\F$-algebras $\textbf{A} = (A_1,\ldots, A_k)$, an
\emph{evaluation} of $f$ in $\textbf{A}$ is given by a $k$-tuple of maps $\boldsymbol{v}=(v_1,v_2,\ldots,v_k)$, where $v_i: \ex_i \to
A_i$. We can extend it to the map $\boldsymbol{v}:
\F\angle{\ex_1}\otimes\cdots\otimes\F\angle{\ex_k}\to A_1\otimes
\cdots\otimes A_k$ as follows: For any monomial $m =
m_1\otimes\cdots\otimes m_k$ where $m_i\in \ex^*_i$, let $\boldsymbol{v}(m) = v_1(m_1)\otimes\cdots\otimes v_k(m_k)$. In particular, for each $x\in \ex_j$ let $\boldsymbol{v}(x) = 1_1
\otimes\cdots\otimes v_j(x)\otimes \cdots\otimes 1_k$ where $1_j$ is the multiplicative identity of $A_j$. We can now extend
$\boldsymbol{v}$ by linearity to all polynomials in the domain $\F\angle{\ex_1}\otimes\cdots\otimes\F\angle{\ex_k}$.

Next, we define a \emph{partial evaluation} of $f\in \F\angle{\ex_1}\otimes\cdots\otimes\F\angle{\ex_k}$ in $\textbf{A}$. Let $k'<k$ and $\boldsymbol{\hat{A}} = (A_1,\ldots, A_{k'})$ be a $k'$-tuple of $\F$-algebras. A partial evaluation of $\F\angle{\ex_1}\otimes\cdots\otimes\F\angle{\ex_k}$ in
$\boldsymbol{\hat{A}}$ is given by a $k'$-tuple of maps $\boldsymbol{\hat{v}} = (v_1 , \ldots, v_{k'})$, where $v_i: \ex_i \to A_i$.  Now, we can define $\boldsymbol{\hat{v}}:
\F\angle{\ex_1}\otimes\cdots\otimes\F\angle{\ex_k}\to A_1\otimes
\cdots\otimes A_{k'}\otimes \F\angle{\ex_{k+1}}\otimes \cdots \otimes
\F\angle{\ex_k}$ as follows.  For a monomial $m =
(m_1\otimes\cdots\otimes m_k)$,  $m_i\in \ex^*_i$, we let $\boldsymbol{\hat{v}}(m) = v_1(m_1)\otimes\cdots\otimes v_{k'}(m_{k'})\otimes m_{k'+1}\otimes\cdots\otimes m_k$. By linearity, the partial evaluation $\boldsymbol{\hat{v}}$ is defined for any $f\in\F\angle{\ex_1}\otimes\cdots\otimes\F\angle{\ex_k}$ where $\boldsymbol{\hat{v}}$ takes values in $A_1\otimes\cdots\otimes A_{k'}\otimes \F\angle{\ex_{k'+1}}\otimes\cdots\F\angle{\ex_k}$.
 
\vspace{0.14 cm} 

Although it is already implicit, we formally recall that  when we consider ABPs over $\F\angle{\ex_1}\otimes\cdots\otimes\F\angle{\ex_k}$ the linear forms are defined over tensors 
of the form $1\otimes \cdots \otimes x_{ij} \otimes \cdots \otimes 1$. These tensors play the role of a variable in the tensor product structure. 

\vspace{0.18 cm}
\noindent\textbf{A few more useful notations :}  
Let $S_{k,{n},d,s}$ denote the set of all polynomials in $\F\angle{\ex_1}\otimes\cdots\otimes\F\angle{\ex_k}$ computed by ABPs of size $s$ and $d$ layers.
Following the notation in
Theorem~\ref{forbesshpilka}, let $\mathcal{H}_{k,{n},d,s}$ be a hitting set for $S_{k,{n},d,s}$. That is, $\mathcal{H}_{k,n,d,s}$ is a
collection of evaluations $\boldsymbol{v} = (v_1,\ldots,v_k)$, 
such that for any nonzero
polynomial $f\in S_{k,{n},d,s}$ there is an evaluation $\boldsymbol{v} =
(v_1,\ldots,v_k)\in \mathcal{H}_{k,{n},d,s}$ such that
$\boldsymbol{v}(f)$ is a nonzero matrix. 
Forbes and Shpilka \cite{FS13} have constructed a quasi-polynomial size hitting set $\mathcal{H}_{1,{n},d,s}$. (see Theorem~\ref{forbesshpilka}).  The following lemma shows an efficient bootstrapped construction of a hitting set $\mathcal{H}_{k,{n},d,s}$ for the set $S_{k,{n},d,s}$ of polynomials, using the hitting set $\mathcal{H}_{1,{n},d,s}$.

More formally, we state the following lemma.
\begin{lemma}\label{pcabppit}
  There is a set of evaluation maps $\mathcal{H}_{k,{n},d,s} = \{(v_1,\ldots,v_k): v_i\in\mathcal{H}_{1,{n},d,s_k}\}$ where $s_k = s(d+1)^{(k-1)}$ such that, for $i\in [k]$, we have $v_i :\ex_i \to \mathcal{M}_{d+1}(\F)$, and 
 $\mathcal{H}_{k,{n},d,s}$ is a hitting set for the class of polynomials $S_{k,{n},d,s}$. Moreover, the size of the set 
 is at most $({n}skd)^{O(k^2\log d)}$, and it can be constructed in deterministic $({n}skd)^{O(k^2\log d)}$ time.
\end{lemma} 
 
Once we prove the above lemma, we will be done with the identity test, since we need to only evaluate the input polynomial on the points 
in the hitting set and check whether the polynomial evaluates to nonzero on any such point.  

Before presenting the proof, we discuss two important ingredients. A
polynomial $f$ in $\F\angle{\ex_1}\otimes\cdots\otimes\F\angle{\ex_k}$ can be written
as $f=\sum_{m\in {X}^*_k} f_m \otimes m$ where each $m$ is a monomial
over variables $\ex_k$ and $f_m \in \F\angle{\ex_1}\otimes\cdots
\otimes\F\angle{\ex_{k-1}}$. Given that $f$ has a small ABP, we first show
that each polynomial $f_m$ also has a small ABP. 
%
%

\begin{lemma}\label{coeff-abp}
  For each $f\in S_{k,{n},d,s}$ and $m\in {\ex}^{*}_k$, the polynomial
  $f_m \in \F\angle{\ex_1}\otimes\cdots \otimes \F\angle{\ex_{k-1}}$
  has an ABP of size $s (d+1)$ and $d$ layers.
\end{lemma}

\begin{proof}
  Suppose $f \in \F\angle{\ex_1}\otimes\cdots \otimes
  \F\angle{\ex_{k}}$ has an ABP $B$ of size $s$ and 
  $m=x_{i_1k} x_{i_2k}\cdots x_{i_{\ell}k}$ where some of the indices could be repeated. We create a copy of $f$ in $\F\angle{\ex}$ where $\ex=\cup_{i=1}^k\ex_i$ in an obvious way: 
  Just substitute $1\otimes\cdots \otimes x_{ij} \otimes \cdots \otimes 1$ terms present on the edge label of the ABP for $f$ by $x_{ij}$. Call this copy $g$ and its ABP $B$ (with a little abuse of notation). 
  Now, we construct an automaton $A$ that isolates precisely those words (monomials) $w\in \ex^*$ from $g$ such that $w\vert_{{X}_k} = m$.  The automaton $A$ is depicted in Figure~\ref{fig1}.

\begin{figure}[h]
\begin{center}
\begin{tikzpicture}[scale=0.7]
\node(pseudo) at (-1,0){};
 \node(0) at (-6,0)[shape=circle,draw] {$q_0$}; 
     \node(1) at (-2.5,0)[shape=circle,draw] {$q_{1}$};
     \node(2) at (.8,0)[shape=circle,draw] {$q_{2}$}; 
     \node(3) at (5,0)[shape=circle,draw,double] {$q_{\ell}$}; 
     \path [->]
     
  (0)      edge [right=27]  node [above]  {$x_{i_1k}$}     (1)
  (1)      edge [right=25]  node [above]  {$x_{i_2k}$}     (2)
  
  (2)      edge [right=25,dotted]  node [above]  {$x_{i_{\ell}k}$}     (3)
  
  (0)      edge [loop above]    node [above]  {$\bigcup_{i=1}^{k-1} \ex_i$}    () 
 (1)      edge [loop above]    node [above]  {$\bigcup_{i=1}^{k-1} \ex_i$}       ()
 (2)      edge [loop above]    node [above]  {$\bigcup_{i=1}^{k-1} \ex_i$}    () 
 (3)      edge [loop above]    node [above]  {$\bigcup_{i=1}^{k-1} \ex_i$}    (); 

\end{tikzpicture}

\caption{The transition diagram of the automaton $A$}\label{fig1}
\end{center} 
\end{figure}
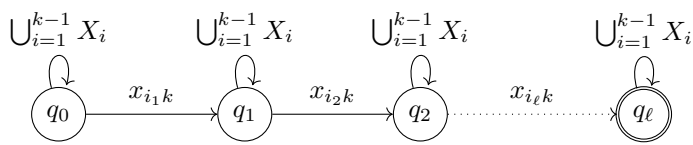

The automaton simply loops around in each state $q_t$ if the input letter is in $\bigcup_{i=1}^{k-1} \ex_i$. It makes a forward transition from $q_t$ to $q_{t+1}$ only on reading $x_{i_{t+1}k}$, for $0\leq t\leq \ell-1$.

Naturally, the ABP $B$ can be thought of as a $\F$-weighted acyclic automaton $B$ without any  $\epsilon$-transition \footnote{In fact any ABP can be also represented by an weighted acyclic automaton of similar size, such that the polynomial computed by the ABP and the finite series computed by the automaton are the same.} computing same $g$. Now we compute the Hadamard product of $B$ with $A$, denoted by $B\odot A$ over the \emph{free} monoid computing $g\odot A$. By a
basic result of Sch\"{u}tzenberger \cite[Theorem 3.2, pp.~428]{Sak09}, it is known that $B\odot A$ has an automaton of size $s (\ell + 1)$. 
This is basically the computation of intersection of two weighted automata and it can be easily observed that the resulting automata is also an ABP. 
\end{proof}

For a polynomial $f$ in $\F\angle{\ex_1}\otimes\cdots \otimes
\F\angle{\ex_{k}}$, consider a partial evaluation $\boldsymbol{v} = (v_1,\ldots,v_{k-1})$ such that each $v_i: \ex_i\to \mathcal{M}_{t_i}(\F)$. The evaluation $\boldsymbol{v}(f)$ is a $T\times
T$ matrix with entries from $\F\angle{\ex_k}$, where $T = t_1t_2\cdots t_{k-1}$.

\begin{lemma}\label{lemma-abpsize}
  For each $p,q\in [T]$, the $(p,q)^{th}$ entry of $\boldsymbol{v}(f)$ can be computed by an ABP of size $sT$ and $d$ layers.
\end{lemma}
The proof is routine and included in 
the appendix. Now we are ready to prove Lemma~\ref{pcabppit}.

\vspace{0.12cm}

\begin{tjproof}
  The proof is by induction on $k$. For the base case $k=1$ the hitting set $\mathcal{H}_{1,{n},d,s}$ from Theorem~\ref{forbesshpilka} suffices. Note that any nonzero $f \in S_{k,{n},d,s}$ can be written as $f=\sum_{m\in {X}^*_k} f_m \otimes m$
where each $m$ is a monomial over $\ex_k$ and $f_m \in
\F\angle{\ex_1}\otimes\cdots \otimes\F\angle{\ex_{k-1}}$. Since
$f\not \equiv 0$ we must have $f_m\not \equiv 0$ for some $m\in \ex^*_k$ . Moreover, by Lemma~\ref{coeff-abp} we know that for each $m\in
{\ex_k}^{*}$ the polynomial $f_m \in \F\angle{\ex_1}\otimes\cdots
\otimes \F\angle{\ex'_{k-1}}$ can be computed by an ABP of size $s (d+1)$. Let $s' = s (d+1)$. 

By the inductive hypothesis $f_m$ evaluates to nonzero on some point in the set :  $\mathcal{H}_{k-1,n,d,s'} = \{ (v_1,v_2,\ldots,v_{k-1}) | v_i \in
\mathcal{H}_{1,{n},d,s'_{k-1} } \}$ where $s'_{k-1} = s' (d+1)^{k-2} = s (d+1)^{k-1}$.  

Hence, there is an evaluation $\boldsymbol{v'} \in \mathcal{H}_{k-1,{n},d,s'}$ such
that $\boldsymbol{v'}(f_m)$ is a nonzero matrix of dimension $(d+1)^{k-1}$.
Interpreting $\boldsymbol{v'}$ as a \emph{partial evaluation} for $f$, we observe that
$\boldsymbol{v'}(f)$ is a $(d+1)^{k-1} \times (d+1)^{k-1}$ matrix with entries from $\F\angle{\ex_k}$. Since $\boldsymbol{v'}(f_m) \neq 0$,
it follows that some $(p,q)^{th}$ entry of $\boldsymbol{v'}(f)$ is a nonzero polynomial $g \in \F\angle{\ex_k}$. By
Lemma~\ref{lemma-abpsize}, each entry of $\boldsymbol{v'}(f)$ has an
ABP of size $s(d+1)^{k-1}$. In particular, $g
\in S_{1,{n},d,s(d+1)^{k-1}}$  and it follows from Theorem~\ref{forbesshpilka} that there is a an evaluation $v''$ in $\mathcal{H}_{1,{n},d,s(d+1)^{k-1}}$  such that ${v''}(g)$ is a nonzero matrix of dimension $(d+1) \times (d+1)$.

Thus, for the combined evaluation map $\boldsymbol{v}=(\boldsymbol{v'},{v''})$, it follows that $\boldsymbol{v}(f)$ is a nonzero matrix of dimension $(d+1)^k
\times (d+1)^k$.  Define $\mathcal{H}_{k,{n},d,s}= \{(v_1,\ldots,v_k): v_i\in
\mathcal{H}_{1,{n},d,s_k}\}$,
where $s_k=s (d+1)^{k-1}$.  However, from the inductive hypothesis, we know that
$\boldsymbol{v'} = (v_1,\ldots,v_{k-1}) \in \mathcal{H}_{k-1,{n},d,s (d+1)}$ where each $v_i\in \mathcal{H}_{1,{n},d,s(d+1)^{k-1}}$.
Therefore,
 $\boldsymbol{v}=(\boldsymbol{v'},{v''}) \in
\mathcal{H}_{k,{n},d,s}$ and $\mathcal{H}_{k,{n},d,s}$ is a
hitting set for the class of polynomials $S_{k,{n},d,s}$.


Finally, note that $|\mathcal{H}_{k,{n},d,s}| =
|\mathcal{H}_{1,{n},d,s_k}|^k$.  Since $
|\mathcal{H}_{1,{n},d,s_k}| \leq ({n}ds_k)^{O(\log d)} $,
it follows that $|\mathcal{H}_{k,{n},d,s}| \leq ({n}skd)^{O(k^2 \log d)}
$. Clearly, the set $\mathcal{H}_{k,{n},d,s}$ can be constructed in the claimed running time.
\end{tjproof}

\section{Randomized Algorithm for Zero Testing of Weighted Automata Over $k$-Monoids}\label{random-pit}

We now consider pc monoids more general than $k$-clique monoids, over which too we can do efficient zero testing of automata. A  \emph{$k$-monoid} is a pc monoid $M$ whose non-commutation graph $G_M$ is a union of subgraphs $G_M=G_1\cup G_2$ such that $G_1$ has a clique edge cover of size $k'$ and $G_2$ has a vertex cover of size $k-k'$. It follows that $G_M$ has a $k$-covering of cliques and star graphs. For the application, we will assume that this $k$-covering of $G_M$ is explicitly given as part of the input. In this section $\F\angle{M}$ is used to denote the $\F$-algebra generated by the monoid $M$. 



\begin{lemma}\label{tensor-combine}
  Let $\{M_i\}_{i=1}^k$ be pc monoids defined over disjoint 
  variable sets $\{\ex_i\}_{i=1}^k$, respectively. For each $i$, suppose $A_i$ is a randomized procedure that outputs an evaluation $v_i:\F\angle{M_i}\to \mathcal{M}_{t_i(d)}(\F)$ such that for any polynomial $g_i$ in $\F\angle{M_i}$ of degree at most $d$, $g_i$ is nonzero if and only if $v_i(g_i)$ is a nonzero matrix with probability at least $1-\frac{1}{2k}$.

Then, for the evaluation  $\boldsymbol{v} : \F\angle{M_1}\otimes\cdots\otimes \F\angle{M_k} \to
\mathcal{M}_{t_1(d)}(\F) \otimes\cdots\otimes \mathcal{M}_{t_k(d)}(\F)
$ such that $\boldsymbol{v} = (v_1,\ldots,v_k)$
and any nonzero polynomial $f \in
\F\angle{M_1}\otimes\cdots\otimes \F\angle{M_k}$ of degree at most $d$,
the matrix $\boldsymbol{v}(f)$ is nonzero with probability at least
$1/2$.
\end{lemma}  

\begin{proof}
The proof is by induction on $k$. For the base case $k=1$, it is trivial.
Let us fix an $f \in \F\angle{M_1}\otimes\cdots\otimes \F\angle{M_k}$ of degree at most $d$ such that $f \not \equiv 0$.
The polynomial $f$ can be written as 
$f=\sum_{m\in \M_k} f_m \otimes m$ where $m$ are the words over the pc monoid $M_k$ and $f_m \in \F\angle{M_1}\otimes\cdots \otimes\F\angle{M_{k-1}}$. Since $f\not\equiv 0$ we must have $f_m\not\equiv 0$ for some $m\in M_k$. 

Now, inductively we have the evaluation $\boldsymbol{v'}  = (v_1,\ldots,v_{k-1})$
 for the class of polynomials in $\F\angle{M_1}\otimes\cdots \otimes\F\angle{M_{k-1}}$ of degree at most $d$. 
 Since $f_m \not \equiv 0$, with high probability $\boldsymbol{v'}(f_m)$ is a nonzero matrix of dimension 
 $\prod_{i=1}^{k-1} t_i(d)$. By induction the failure probability is bounded by $\frac{k-1}{2k}$. 
 
As $\boldsymbol{v'}$ is a \emph{partial evaluation} for $f$, we observe that $\boldsymbol{v'}(f)$ is a  matrix  of dimension $\prod_{i=1}^{k-1} t_i(d)$ whose entries are polynomials in $\F\angle{M_k}$.
Since $\boldsymbol{v'}(f_m) \neq 0$ we conclude that some $(p,q)^{th}$ entry of $\boldsymbol{v'}(f)$ contains a nonzero polynomial $g \in \F\angle{M_k}$ of degree at most $d$. 
Choose the evaluation $v_k\in S_k$ which is the output of the randomized procedure $A_k$, such that $v_k(g)$ is a nonzero matrix of dimension $t_k(d)$.
Hence, for the combined evaluation $\boldsymbol{v}=(\boldsymbol{v'},v_k)$,  $\boldsymbol{v}(f)$ is a nonzero matrix of dimension $\prod_{i=1}^{k} t_i(d)$. Using an union bound the failure probability can be bounded by $1/2$.  
\end{proof}


For the proof of Theorem \ref{main-theorem-2}, we first give a randomized
polynomial-time identity testing algorithm for polynomials over pc monoids whose non-commutation graph is a star graph.
  
\begin{lemma}\label{star-cover}
  Let $M = ((\ex\cup y)^*,I)$ be a monoid whose non-commutation graph $G_M$ is a star graph with center $y$. Then
  for any constant $k$, there is a randomized procedure that outputs
  an evaluation $v: \ex\cup\{y\} \to \M_{t(d)}(\F)$ where $t(d)$ is at
  most $d$, such that for any polynomial $f\in \F\angle{M}$ of degree
  at most $d$, the polynomial $f$ is nonzero if and only if $v(f)$ is
  a nonzero matrix. The success probability of the algorithm is at
  least $1-\frac{1}{2k}$.
\end{lemma}

\begin{proof}
  If $f$ is nonzero, then there exists a monomial $m$ in $M$ with
  nonzero coefficient. The idea is to isolate all monomials in
  $\{\ex\cup y\}^*$ that are equivalent to $m$ in $M$. Let the degree
  of $y$ in monomial $m$ be $\ell\leq d$. Then $m$ can be written as
  $m = m_1 y m_2 \cdots m_{\ell} y m_{\ell+1}$ where each $m_i$ is a
  word in $\ex^*$. As $\ex$ is a commuting set of variables, any
  permutation of $m_i$ produces a monomial equivalent to $m$ in $M$.
  Now consider the automaton in Figure~\ref{fig2}.

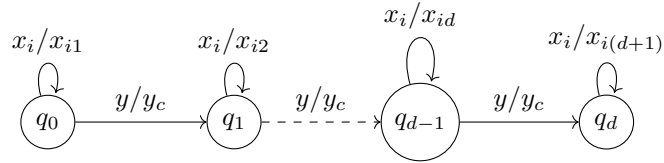
\begin{figure}[h]
\begin{center}
\begin{tikzpicture}[scale=0.7]
\node(pseudo) at (-1,0){};
 \node(0) at (-6,0)[shape=circle,draw] {$q_0$}; 
     \node(1) at (-2.5,0)[shape=circle,draw] {$q_{1}$};
     \node(2) at (1,0)[shape=circle,draw] {$\small q_{d-1}$}; 
    \node(3) at (4.5,0)[shape=circle,draw] {$q_{d}$}; 
     \path [->]
     
  (0)      edge [right=27]  node [above]  {$y/y_c$}     (1)
  (1)      edge [right=25,dashed]  node [above]  {$y/y_c$}     (2)
 
  (2)      edge [right=25]  node [above]  {$y/y_c$}     (3)
  
  (0)      edge [loop above]    node [above]  {$x_i / x_{i1}$}    () 
 (1)      edge [loop above]    node [above]  {$x_i / x_{i2}$}       ()
 (2)      edge [loop above]    node [above]  {$x_i /x_{id}$}    () 
 (3)      edge [loop above]    node [above]  {$x_i / x_{i(d+1)}$}    (); 

\end{tikzpicture}

\caption{The transition diagram of the automaton}\label{fig2}
\end{center} 
\end{figure}

Let $m$ as $m = m_1 y m_2 \cdots m_{\ell} y m_{\ell+1}$, where each $m_i$ is a
maximal substring of $m$ in $\ex^*$. We refer to the $m_i$ as blocks. 
The above automaton 
 keeps count of blocks as it scans the monomial $m$. As it scans $m$, if the automaton is in the $j^{th}$ block, it substitutes each variable $x_i\in\ex$ read by a corresponding commuting variable $x_{ij}$ where the index $j$ encodes the block number. The $y$ variable is renamed by a commutative variable $y_c$. 
In effect, we substitute each $x_i$ and $y$ by the 
transition matrices $N_{x_i}$ and $N_y$ of dimension $d+1$. The transition matrices are explicitly given below. 
\[
N_{x_i} = 
\begin{bmatrix}
x_{i1} &0 &0 &\ldots &0\\
0 &x_{i2} &0 &\ldots &0\\
\vdots &\vdots &\ddots &\ddots &\vdots\\
0 &0 &\ldots &x_{id} &0 \\
0 &0 &\ldots &0 &x_{i(d+1)}\\
\end{bmatrix},
\quad\quad
N_{y} = 
\begin{bmatrix}
0 &y_c &0 &\ldots &0\\
0 &0 &y_c  &\ldots &0\\
\vdots &\vdots &\ddots &\ddots &\vdots\\
0 &0 &\ldots &0 &y_c\\
0 &0 &\ldots &0 &0\\
\end{bmatrix}.
\]



Now we explain this matrix substitution. Let $f=\sum_m \alpha_m m$, where $\alpha_m\in\F$. 
We write $f=\sum_{\ell=1}^d f_\ell$,
where $f_\ell=\sum_{m:\deg_y(m)=\ell}\alpha_m m$. That is, $f_\ell$ is the
part of $f$ consisting of monomials $m$ with $y$-degree $\deg_y(m)=\ell$.

From the description of the automaton, we can see that for each $\ell\in[d]$, the $(0,\ell)^{th}$ entry of the output matrix is the
commutative polynomial $f^{c}_{\ell}\in\F[\{x_{i,j}\}_{1\leq i\leq n, 1\leq j\leq d+1}, y_c]$. The construction ensures the following.

\begin{observation}\label{crucial-obs}
For each $0\leq \ell \leq d$, $f_{\ell}=0$ if and only if $f^{c}_{\ell}=0$. 
\end{observation}
 
The randomized identity test is by substituting random scalar values for the commuting variables $x_{ij}$ and $y_c$ from a set $S\subseteq \F$ of size at least $2kd$, such that the output matrix becomes nonzero. The bound on the success probability follows from Polynomial
Identity Lemma \cite{Zip79, Sch80, DL78}.
\end{proof}
    \vspace{-0.25 cm}

Now we are ready to prove Theorem \ref{main-theorem-2}.   
 \vspace{-0.3 cm}
\begin{proof}
  Let $M'$ be a pc monoid whose non-commutation
  graph $G_{M'}$ is a clique. Let $g\in\F\angle{M'}$ be a nonzero
  polynomial of degree at most $d$. By the Amitsur-Levitzki Theorem~\cite{AL50}, if we substitute variables $x_i\in M'$ by generic matrix of size $d$ over the variables $\{x^{(i)}_{u,v}\}_{1\leq u, v\leq d}$, the output matrix is nonzero.\footnote{In fact the Amitsur-Levitzki theorem guarantees that generic matrices of size $\lceil{\frac{d}{2}\rceil} + 1$ suffice \cite{AL50}.} Moreover, the entries of the output matrix are commutative
  polynomials of degree at most $d$ in the variables
  $\{x^{(i)}_{u,v}\}_{1\leq i\leq n, 1\leq u, v\leq d}$. It suffices to
 randomly substitute for each $x^{(i)}_{u,v}$ variable from a set $S\subseteq \F$ of size at least $2kd$. This defines the evaluation map $v : \F\angle{M'}\rightarrow \mathbb{M}_d(\F)$. The resulting identity test  succeeds with probability at least $1-\frac{1}{2k}$. For the star graphs, the
  evaluation map is already defined in Lemma~\ref{star-cover}. 
  
  Given a $\F$-weighted automaton $A$ of size $s$ over a $k$-monoid $M = (\ex^*,I)$, by Theorem~\ref{equivalence-to-pit}, the zero testing of $A$ reduces to identity testing of a collection of  ABPs of the form :  $f=u^{T} N^{d} v$ over $\F\angle{M}$, where $N$ is the transition matrix of $A$ and $d$ is bounded by $O(n^3 s^2)$. 
  Now, to test identity of $f$ where $M$ is a $k$-monoid, it suffices to test identity of $\psi(f)$ where $\psi$ is the injective homomorphism from Lemma \ref{main-iso-lemma}. 
  Now $\psi(f)$ in $\F\angle{M'_1}\otimes\cdots\otimes \F\angle{M'_k}$, where for each $i\in [k]$
  the non-commutation graph of $M'_i$ is either a clique or a star. 
  
  By Lemma~\ref{tensor-combine}, we can construct the 
  evaluation map $\boldsymbol{v}=v_1\otimes v_2\otimes\cdots\otimes
  v_k$ where for each $i\in [k]$, $v_i$ is an evaluation map for either
  a clique or a star graph depending on $M'_i$. The range of $\boldsymbol{v}$ is matrices of dimension at most
  $d^k$, which is bounded by $(sn)^{O(k)}$ as $d$ is bounded by $O(n^3 s^2)$.
  This completes the proof of Theorem~\ref{main-theorem-2}.
\end{proof} 

\section*{Acknowledgement} 
We thank the anonymous reviewers for their invaluable feedback that helped on improving the manuscript. In particular, we are grateful to an anonymous reviewer for suggesting simplified proofs for some lemmas, the concept of valid automata, and pointing out the earlier results related to partially commutative monoids which are relevant to the current work. We also appreciate their comments on mathematical terminology consistent with weighted automata literature.  

\bibliography{ref2}

\appendix

\section{The Proof of Lemma \ref{main-iso-lemma}}
\begin{proof}
It is straightforward to check that $\psi$ is a ring
homomorphism.
To show the injectivity, it is enough to show that $\psi(m) = \psi(m')$ implies $m = m'$ in $M$ for
any words $m,m'\in M$. 
We prove the claim by induction on the length of words in $M$. Suppose that
for words $m\in M$ of length at most $\ell$, if $m'$ is not
$\tilde{I}$-equivalent to $m$ then $\psi(m)\ne \psi(m')$. The
base case, for $\ell=0$ clearly holds.

Now, suppose $m = x\cdot m_1\in\ex^{\ell+1}$ for $x\in\ex$ and
$\psi(m)=\psi(m')$.

\begin{claim}
For some $m_2\in M$, $m' = x\cdot m_2$ in $M$.
\end{claim}
\begin{proof}
Assume, to the contrary, that there is no $m_2\in M$ such that $m' =
x\cdot m_2$.  Let $J=\{j\in[k]\mid x\in\ex_j\}$. If the variable $x$
does not occur in $m'$ then $m\vert_{\ex_j}\neq m'\vert_{\ex_j}$ for each $j\in J$. This
implies that $\psi(m) \neq \psi(m')$ which is a contradiction.

On other hand, suppose $x$ occurs in $m'$ and it cannot be moved to
the leftmost position in $m'$ using the commutation relations in $I$.
Then we must have $m' = ayxb$ for some $y\in\ex_j$ and $j\in J$, where
$a,b\in \ex^*$, for the leftmost occurrence of $x$ in $m'$. Hence
$m\vert_{\ex_j}\neq m'\vert_{\ex_j}$, because $x$ is the first
variable in $m\vert_{\ex_j}$ and $x$ comes after $y$ in
$m'\vert_{\ex_j}$.  Therefore, $\psi(m) \neq \psi(m')$ which is a
contradiction.
\end{proof} 

Now,  $\psi(x\cdot m_1) = \psi(x\cdot m_2)$ implies that
$\psi(m_1) = \psi(m_2)$. Both $m_1$ and $m_2$ are of length $\ell$. By induction 
hypothesis it follows that $m_1 = m_2$, and hence $m = m'$.
\end{proof}

\section{The Proof of Lemma \ref{lemma-abpsize}}
\begin{proof}

In effect the edges of the input branching program $B$ are now labelled by matrices of dimension $T$ with entries are linear forms over the variables $\ex'_k$. To show that each entry of the final $T\times T$ matrix can be computed by an ABP of size $sT$, let us fix some $(i,j)$ such that $1\leq i,j\leq T$ and construct an ABP $B'_{ij}$ computing the polynomial in the $(i,j)^{th}$ entry.

The construction of $B'_{ij}$ is as follows. We make $T$ copies of each node $p$ (except the source and sink node) of $B$ and label it as $(p,k)$ for each $k\in [T]$. Let us fix two nodes $p$ and $q$ from $B$ such that there is a $T\times T$ matrix $M_{pq}$ labelling the edge $(p,q)$ after the substitution. Then, for each $j_1,j_2\in [T]$, add an edge between $(p,j_1)$ and $(q,j_2)$ in $B'_{ij}$ and label it by the $(j_1,j_2)^{th}$ entry of $M_{pq}$.
When $p$ is the source node, for each $j_2\in T$, add an edge between the source node and $(q,j_2)$ in $B'_{ij}$ and label it by the $(i,j_2)^{th}$ entry of $M_{pq}$. Similarly, when $q$ is the sink node, for each $j_1\in T$, add an edge between $(p,j_1)$ and the sink node in $B'_{ij}$ and label it by the $(j_1,j)^{th}$ entry of $M_{pq}$. 

We just need to argue that the intermediate edge connections simulate matrix multiplications correctly. This is simple to observe, since for each path $\mathcal{P} = \{(s,p_1),(p_1,p_2),\ldots,(p_{\ell - 1},t)\}$ in $B$ (where $s,t$ are the source and sink nodes respectively) and each $(j_1,\ldots,j_{\ell - 1})$ such that $1\leq j_1,\ldots,j_{\ell - 1} \leq T$, there is a path $(s,(p_1,j_1)), ((p_1,j_1),(p_2,j_2)),\ldots, ((p_{\ell-1}, j_{\ell-1}),t)$ in $B'_{ij}$ that computes $M_{(s,p_1)}[i,j_1]M_{(p_1,p_2)}[j_1,j_2]\cdots M_{p_{\ell-1},t}[j_{\ell-1},j]$ where $M_{(p,q)}$ is the $T\times T$ matrix labelling the edge $(p,q)$ in $B$.
The size of $B'_{ij}$ is $sT$, and the number of layers is $d$. 
\end{proof}
\end{document}